# A Quasi-experimental Study of Conversational AI-supported Engagement in Large CS Courses

Neha Rani [1,*], Sharan Majumder [1], Ishan Bhardwaj [1], and Pedro Guillermo Feijóo-García [2]

[1] Department of Computer and Information Science and Engineering, University of Florida, Gainesville, USA
[2] School of Computing Instruction, College of Computing, Georgia Institute of Technology, Atlanta, USA
Email: neharani@ufl.edu (N.R.); smajumder1@ufl.edu (S.M.); ishanbhardwaj@ufl.edu (I.B.); pfeijoogarcia@gatech.edu (P.G.F.-G.)
*Corresponding author

***Abstract*—Lucrative career prospects and creative opportunities often attract students to enroll in computer science courses, resulting in large class sizes. A common challenge in large classrooms is the lack of engagement between students and both the instructor and the learning material. Recent advances in Large Language Models (LLMs) present an opportunity to leverage Conversational Artificial Intelligence (CAI) to address this challenge. To examine the potential of CAI to support engagement with learning material, we designed an in-class activity in a large Software Engineering course in which students interacted with a CAI tool. Using a quasi-experimental, within-subject study design in a real classroom at a large public United States University. We compared student engagement during an activity that involved the use of CAI for discussion with an activity that did not (peer discussion). For our analysis, we collected data on students' interactions with the CAI tool and their self-reported responses on multiple engagement scales. Results show higher student engagement with the learning material in the activity involving CAI discussion as opposed to without (peer discussion). Higher self-reported learning was also observed in the activity with CAI as compared to without. Students also reported significantly more usefulness for learning in the CAI condition. Overall, our results indicate that CAI (ChatGPT) has the potential to support engagement with learning content during in-class activities in large class sizes.**

## I. INTRODUCTION

Over the past decade, the advancement in Machine Learning (ML) and Large Language Models (LLMs) have revolutionized Artificial Intelligence (AI) technology and its applications. LLM-based Conversational AI (CAI) applications are now able to interpret and understand queries and produce intelligent responses with more and more accuracy. Given the accuracy of responses, level of detail, and human-like narration, AI tools are finding applications in various aspects of education, both for learners and educators [1–4]. Many view AI as a pivotal force that could potentially spark a transformative shift in education [5]. The past decade has witnessed extensive research on AI in education [4]. The exploration of possibilities of using CAI in education particularly exploded with the introduction and public availability of OpenAI's ChatGPT in November 2022 [6]. CAI tools have been used to provide support in various aspects of informal and formal education [6, 7]. CAI has been used extensively to assist in learning programming by supporting source code generation, explaining programming errors, and providing solutions [8]. These CAIs are envisioned to revolutionize pair programming. It is reintroduced as a collaborative coding effort in which programmers and AI work together [9]. While these are avenues for the formal use of AI, the informal use of AI is more diverse and rampant due to AI tools being freely available [7]. Even with AI's shortcomings in Education, such as plagiarism, there are many avenues in which AI assistance is valuable. AI has the potential to assist in the learning process by providing feedback, interactive discussion, clarifying doubts, recommending learning content, and so on [10, 11]. AI can be employed to support the learning process in multiple other innovative ways. Supporting interaction and engagement with learning content is one such avenue through which AI can support education [12].

Computer science courses are observing a sharp increase in enrollment, with large classrooms becoming a common phenomenon. This increase in enrollment can be attributed to lucrative technology career pathways. Lucrative opportunities attract even students with non-computer science majors to enroll in computer science courses. Limitations in staffing and resources in modern higher education have also contributed to the emergence of large class sizes [13]. With hundreds of students in the classroom, it is particularly challenging for instructors to maintain engagement and one-to-one interactions with students. Previous case studies have found that a class size of just over 45 is too large for effective learning and teaching [14, 15]. Even with different student engagement techniques, such as in-class discussions, in-class activities, and other active learning strategies, it is difficult to involve a large number of students due to time constraints and syllabus completion. Lower engagement in large classes can also be attributed to hesitation and fear of speaking up and/or being judged by a large number of peers. Nevertheless, it is critical for students to stay engaged, particularly with learning content, to reach their full academic potential [16]. Therefore, exploring means to support student engagement is of utmost importance.

A range of pedagogical strategies have been proposed to support student engagement in large classroom settings, including techniques such as think-pair-share and peer discussion [17, 18]. Think-pair-share has been shown to yield better learning outcomes compared to lecture-based learning [18]. Nevertheless, the effectiveness of such methods is not universal and is often constrained by practical challenges, including the logistical complexity of group formation and the lack of initiative by students in starting substantive dialogue. These limitations are particularly pronounced in hybrid learning environments, which have become increasingly prevalent in the post-COVID educational landscape [19]. Alternative teaching approaches, such as Application, Response, Collaboration (ARC), may encourage student participation and engagement by fostering a positive learning environment [20]. However, even with these practices, a significant improvement in engagement is not seen in many classrooms. This could be due to differences between real-life and experimental settings. Moreover, peer-based engagement strategies are inherently limited by the relatively similar knowledge levels among students, which may restrict the informativeness and quality of peer interactions. Some non-AI technological solutions have also been introduced, such as electronic student response systems (commonly referred to as "clickers"), through which anonymous student responses can be submitted and immediate feedback can be received [21]. While these tools facilitate some interaction, they primarily monitor attentiveness and do not substantially promote critical thinking or deeper engagement. The challenge of low engagement is further exacerbated in hybrid classroom formats. Even though students often express a preference for in-person instruction, hybrid modalities are frequently favored for their flexibility, contributing to their widespread adoption [22]. The hybrid format introduces additional complexities in implementing interactive pedagogical methods, as instructors must navigate fluctuating attendance patterns (some students participating online and others in person), variable group dynamics, and limited oversight of individual group interactions. These factors collectively hinder the creation of engaging learning experiences.

We posit that artificial intelligence, particularly LLMs such as ChatGPT, holds significant promise in addressing the persistent challenge of lower student engagement in large-scale higher education classrooms. Contemporary LLM-powered conversational agents exhibit a high degree of intelligence, responsiveness, and personalization, enabling inquiry-based learning that is tailored to individual student needs. Especially with recent advances in LLMs' capacity to produce accurate and humanlike responses, they present an opportunity. AI applications, such as ChatGPT, are intelligent, responsive, accurate, and personalized to a level that inquiry-based learning can help students engage with learning content. Such tools can enable students to engage in more in-depth and thought-provoking conversations with a knowledgeable agent that provides responses tailored to their specific queries, helping them engage more with the learning material. Recent work in this relatively new area has shown promising results in utilizing AI in educational settings. Researchers have proposed that by using AI, students gain opportunities to engage in collaborative learning, and at the same time, teachers could be relieved of duties such as designing assignment problems and administrative tasks [3]. CAI-based tools such as CodeHelp and Code Aid have been explored for learning programming and have been shown to be useful for learning programming [23, 24]. These tools interact with students during programming and help them navigate through the errors they make without providing direct answers.

Prior research has investigated the use of AI for educational purposes. A review of AI in education by Zhai *et al.* [4] reveal that multiple applications of AI have been explored in education for feedback, recommendation, immersive learning, affective computing, and adaptive learning. Generative AI (GAI) is shown to support the creation of learning assignments and personalized learning content [25, 26]. Prior work has shown that the use of AI-driven design for Intelligent Tutoring Systems (ITS) that provide personalized learning environments results in better student engagement [27]. GAI has been used to create personalized exercises and has been found to be useful and engaging [26]. Another investigation explored AI-based learning content generation and learning pathway augmentation [2]. It was found to support learner engagement in an online learning environment. Another prior researcher explored the use of ChatGPT to support engagement in learning. They found positive results; however, the study relied solely on user survey data, which may not reflect real classroom settings [28]. Much of the use of AI to support student engagement research is perception-based [28], in constrained settings, informal learning [29], and limited to a specific population [30]. There is a lack of investigation into the use of AI tools in a real classroom setup to support engagement with learning content in higher education. Based on directions provided by previous research on CAI in education and the current need to support engagement in a real classroom, we conducted a quasi-experimental study to investigate whether the use of AI tools can support engagement in a class activity in a large computer science course. CAI tools such as ChatGPT have the potential to increase engagement by providing immediate, informative responses to students' queries, thus acting as a personalized tutor [12]. We particularly focus on cognitive engagement with learning content [12]. Cognitive engagement involves critical thinking, reflection, exchange of feedback, developing understanding, and so on [12].

A classroom with high engagement is crucial, as it can lead to a better understanding of learning content and, subsequently, result in better performance and appreciation of the course [31]. When students engage with their peers and professors regarding learning material, they begin to thrive and develop an interest in the classroom. Therefore, the issue of low engagement in large classrooms warrants attention. Our research aims to explore whether the use of conversational AI in classroom activities can support students' engagement with learning content in the class. The main research questions we aim to answer through our investigation are:

**RQ1**: Does the use of an AI tool in the classroom activity significantly impact student engagement as opposed to without an AI tool?

**RQ2**: Does the use of an AI tool in a classroom activity impact a student's perception of the usefulness of the learning activity as opposed to one without AI?

To answer these research questions, we conducted a within-subject study using a quasi-experimental research design. The study was conducted in an authentic educational setting, i.e., in an undergraduate computer science core course at a large public university in the United States. The study data were collected during two 50-minute class periods. These two class periods focused on an in-class reading based activity. This activity was a natural part of the course and was not developed for the study. There were approximately 450 students enrolled in the course, of whom 227 volunteered to participate in the study. After filtering the data, 101 participants' data were used for analysis. The study session consisted of students participating in an in-class activity, which involved reading a research article and then having a discussion based on their reading. There were two study conditions: the "AI condition"—in this condition, after reading the article, students had a discussion session with ChatGPT; and the "without AI condition"—in this condition, after reading the article, students had a discussion session with their peers. We discuss our study design in detail in the study description section. Overall, our work explores the potential of Conversational AI to improve student interaction and in-depth engagement with learning content in the classroom.

## II. Background

### A. *Engagement in Education*

Multiple definitions of student engagement are observed in the literature. Overall, student engagement highlights the significance of students being actively involved with the learning activities and interacting with the instructor or learning material [32]. Engagement in a classroom can be broadly classified into three types: behavioral engagement (it relates to learning habits, behaviors, attendance, and so on), social engagement (it refers to the learner responsiveness, their interaction with peers and instructor), and cognitive engagement (it refers to the critical thinking, exploration, deeper understanding of the learning content) [12, 33].

Students' responsiveness and active engagement with the course material, peers, and instructor are instrumental to their learning [34]. This makes it crucial to have better engagement in higher education classrooms, especially large classrooms where a larger number of students are involved. Despite the continued efforts of educators and education researchers, supporting engagement in the classroom remains a challenge. The level of engagement varies between small and large classrooms, with large classrooms often lacking engagement due to logistical and time constraints [35]. With large sizes becoming a common phenomenon, massification is spreading through large universities in higher education [36]. Carini *et al.* [37] report that engagement can be positively related to desirable learning outcomes such as grades and critical thinking. Further, they also revealed that students with the lowest ability benefited most from engagement compared to other students. Therefore, a need is identified to support better engagement in large classes.

National Survey of Student Engagement (NSSE) [38] points out that the core of engagement is to support better academic performance. Engagement is considered to be a predictor of learning and personal development [37]. NSSE lists active, collaborative learning and enriching learning experiences as one of the benchmark factors for their scale, measuring student engagement. It is crucial at this point to differentiate mere responsiveness from engagement. Often, participation is assumed to be synonymous with engagement. Student engagement is a multifaceted concept that plays a pivotal role in education. High levels of engagement are associated with better understanding, retention, and application of knowledge, making it a critical objective for educators [37]. Prior work has shown that including activities that promote student engagement increases attendance and improves academic performance [31]. We, therefore, explore supporting engagement through the use of AI in an in-class activity to generate higher engagement with learning content.

### B. *AI in Education*

With recent technological advances in big data and machine learning, AI has become very powerful. Post the introduction of publicly available tools such as ChatGPT, Claudia, Gemini, and Copilot, the rampant use of AI in everyday life is observed. Artificial Intelligence has emerged as a transformative force across various industries, and education is no exception. The integration of AI into educational environments has the potential to revolutionize traditional teaching and learning processes, offering new ways to enhance student engagement, personalize learning experiences, and improve educational outcomes.

AI in education has been used in domains such as recommendation, feedback, gamification, and adaptive learning. A recent review of the use of ChatGPT in education does a Strengths, Weaknesses, Opportunities, and Threats (SWOT) analysis [12]. They provide a comprehensive report on the use of ChatGPT in education after its release. They argue that the tool has the potential to support education, but there are some challenges as well, so educators may use the tool with caution. AI has revolutionized the field of education in many ways and is even being applied in classrooms. Educators use AI for assistance [39]. AI for educator training is available in many major institutions to promote the use of AI to assist in instruction. A driving force behind adopting AI in education is the surge in students from various locations enrolling in higher education courses offered at a central location, combined with restricted funding [3]. One of the key benefits of AI in education is personalized learning, which can enhance student performance by allowing learners to study at their preferred pace and in a manner that aligns with their individual learning style [40]. Students have also been observed using AI informally as a personal tutor, relying on it for tailored explanations and quick summaries of complex information. Broadly, there are three categories of using AI in education, namely automation of administrative tasks (AI used to perform repetitive tasks by instructors), generating educational content (AI used to generate, synthesize, or summarize learning content), and intelligent tutoring systems (AI used for an intelligent and personalized tutoring system) [3].

## III. Related Works

The popularity of AI has prompted numerous studies aimed at understanding how it can enhance learning within classroom environments [4, 41]. The multifaceted uses of AI in education have been explored, ranging from AI-assisted learning for students to AI-assisted instruction for instructors. Diwanji *et al.* [42] proposed using AI and analytics as a means to support prior learning preparation for flipped classrooms. They point out that preparation prior to the class is challenging for the students, and AI can be leveraged for this purpose. Another popular avenue for using AI applications in education is to build personalized learning assistants. AI-based personalized tutors can support learning in everyday settings [43]. AI assistants have also been explored for higher education classrooms where a low teacher-to-student ratio results in low student-teacher interaction [44]. The use of AI for pedagogical purposes expands even more. AI-powered programming assistant tools such as CodeHelp and CodeAid have been developed and explored to support learning programming in introductory programming courses [23, 24]. So far, they have shown positive results. AI has been explored as a tool to support foreign language learning [45]. AI chatbots for supporting learning have demonstrated the potential to help students learn content in a responsive, interactive, and confidential manner.

AI has already been implemented and deployed for wider use at some educational institutions. For instance, Georgia State University implemented an AI-based chatbot for first-year students taking math and English courses [46]. The chatbot interacted with freshmen 185,000 times and was used by 40,000 students by 2024, and has proved to be useful. In a recent study, a Taiwanese university conducted a study to test the effectiveness of AI chatbots [47]. They found that the academic performance of the experimental group, which used AI to study for an exam, exceeded that of the control group by 8.93%. In terms of self-efficacy, learning attitude, and learning motivation, the experimental group performed better by a medium to large margin [47]. Furthermore, Hadian *et al.* [48] explored the combination of AI and gamification as a novel learning approach. In their investigation, the AI treatment group experienced notable improvements in discussion participation, material access frequency, and academic performance.

There have been a few explorations of the use of AI to support engagement during in-class activities for large computer science classrooms. Adarkwah *et al.* [49] explored AI, specifically ChatGPT, for inquiry-based learning as information-seeking behavior that may support classroom engagement. However, this work was limited to theoretically proposing the use of ChatGPT as a means to support engagement, and no evaluation was conducted. In 2022, Diwan *et al.* [2] investigated the use of AI for learning content and pathways generation to support learner engagement. They explored AI-based narrative fragment generation to create interactive learning narratives. This investigation focused on the development and evaluation of AI systems. This work did not investigate the association between system use and engagement or exploration of the system in real classroom settings. In 2023, another research work proposed combining AI and gamification to support engagement and attendance. This was a position paper and lacked empirical evidence. A recent review investigation by Heung and Chiu [41] reveals multiple studies examining ChatGPT in education to support behavioral, cognitive, and emotional engagement. Their investigation reveals a medium-sized effect on overall student engagement. Prior work has also cautioned the use of generative AI and potential negative impacts, such as over-reliance [1, 50]. Nonetheless, the potential of AI to support education outweighs its potential negative impact and calls for the use of AI with guardrails or supervision. Prior work has highlighted the potential of AI to enhance student engagement and improve learning outcomes [41]. There remains a need to evaluate the role of AI in supporting engagement in authentic classroom settings, especially large classrooms that face more challenges in student engagement. There is also a need to understand how students interact with and respond to AI in formal education settings. Our research aims to address these gaps by conducting both a quantitative and qualitative analysis of interactions between students and AI.

## IV. Study Description

### A. Study Methodology

The study was conducted in an authentic classroom setting of an undergraduate core computer science course. This study was approved by the university's Institutional Review Board exempt process (Protocol: ET00023683). The study was a within-subject design. The study was conducted in two separate 50-minute sessions. The sessions were separated by a week. The study was quasi-experimental in nature as it was conducted in a real classroom. The course in which the study was conducted offered a hybrid setup, allowing students to join the class in person or online via a video conferencing platform, Zoom. The course classes were held on Monday, Wednesday, and Friday. The Friday class is usually an in-class activity, as shown in Fig. 1. The study sessions were conducted during the Friday class as an in-class activity. The usual in-person attendance for the course is below 50% as the class offered a hybrid format. The majority of students joined the class via Zoom. To maintain uniformity and better logistic management for pair formation during study sessions, the class was online only on the days (two Friday classes, Session 1 and Session 2) when the study was conducted. However, it does not significantly affect the authenticity of the classroom environment, as most students were already joining the class online. In addition, the study was conducted during a regular class period using the regular class Zoom.

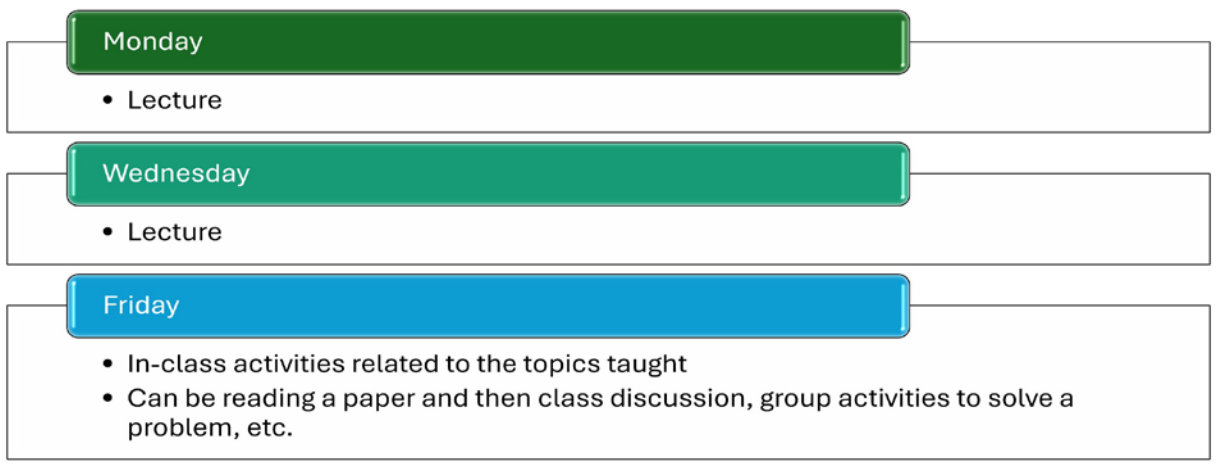

Fig. 1. The usual class plan. The classes were held on Monday, Wednesday, and Friday. Fridays are in-class activities. The study sessions were conducted on Friday, so it seemed like a regular in-class activity.

All the students attending the class participated in the study session. All the students first participated in study Condition 1 (discussion with CAI) and then, after a week, in study Condition 2 (discussion with a peer). The study conditions were not counterbalanced due to the constraints of being part of the actual course class. Ethically, it is not appropriate to treat a group of students differently in the same class. Such an investigation without counterbalance study conditions due to real classroom settings has also been done in prior literature and published [51]. Additionally, the one-week gap between the two conditions reduced the bias of prior condition participation, due to memory loss over time. To track the students' responses while maintaining anonymity, a Participation Identification Number (PID) was randomly generated during the first session, where the students saved that PID to be reused in the next session. In the first session, students experienced a classroom activity that involved the use of the conversational AI tool, ChatGPT. In the second session, students experienced a classroom activity that involved peers and did not involve the use of the conversational AI tool. Students were asked to set up a ChatGPT account for the first study session.

### *B. Study Procedure*

The study consisted of two sessions. The study procedure is illustrated in Fig. 2. During the first session, which involved discussion with a CAI tool, the following steps were taken. First, participants' consent was obtained. Students then completed a pre-study survey about their preferences regarding AI. They were then given 15 min to read an article about REACT, a JavaScript programming Library used to build user interfaces. After this, they spent 10 min interacting with ChatGPT about questions relating to the article. To make sure everyone is interacting with ChatGPT in a similar way, the students were instructed to enter the following prompt that was given to them: "Pretend like you are a troubled student who majors in computer science and has some common questions regarding the REACT framework. Ask me a question, and after I respond, explain to me if I was right or wrong". Subsequently, they copied and pasted the REACT article contents into ChatGPT, and ChatGPT would respond with questions for the student to answer. After the student responds, ChatGPT provides feedback, and then the student asks a follow-up question if they have any or prompts ChatGPT to ask another question. At the end of the discussion, the students would save the ChatGPT conversation link and upload it to the post-study survey, where they would answer questions about their experience interacting with ChatGPT.

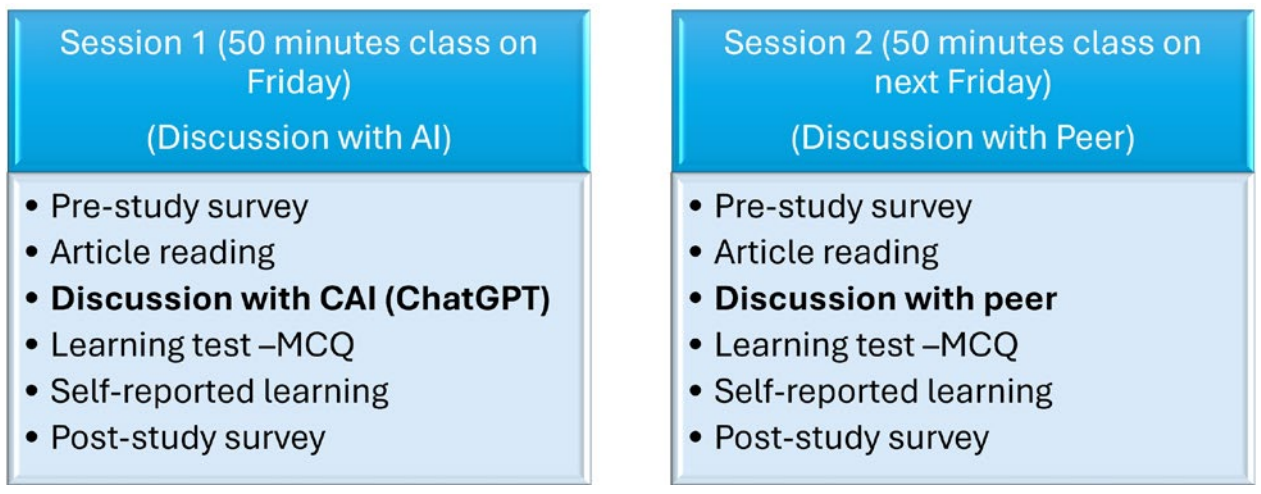

Fig. 2. The two study sessions. On the left is the first session (discussion with Conversational AI condition). On the right is the second session (discussion with peer condition).

In the second session, students performed the classroom activity involving discussion with a peer (without the conversational AI tool). Similar to the first session, they filled out a pre-study survey and were given 15 min to read an article about Angular, a TypeScript-based single-page web application framework. Students were paired randomly with each other and were given 10 min to discuss the article, where they asked and answered questions to each other. In this condition, Student 1 asks a question to Student 2. Student 2 responds, and Student 1 provides feedback. Then Student 2 asks a question, Student 1 responds, and Student 2 provides feedback. Afterward, they save their conversation with their partner and fill out a post-study survey about their experience with the peer discussion activity. It should be noted that the course had provision for an in-class activity once every week. So, the in-class activity used for the study data collection seemed to be part of a regular course activity and was not perceived as an outside course activity by the students.

### *C. Data Collection*

In the first session before the start of the activity, a pre-study survey was conducted, focused on participants' demographics, their prior AI experience, and their AI use expertise level. It also included questions about their proficiency in the reading topic. The post-study survey, deployed at the end of both the study sessions (in-class activity), consisted of multiple-choice questions about the topic in the reading and discussion (peer or CAI) activity, and the usefulness of the discussion (peer or CAI). Two scales were used to measure students' engagement: Student Engagement as a Factor Scale [52] and Student Perspective of Engagement in Class Scale (Assessing Student Perspective of Engagement in Class Tool: ASPECT) [53]. The questions were adapted according to the study. The post-study questionnaire also included 5 Multiple Choice Questions (MCQ) based on the reading topic to assess the level of learning in the activity.

### *D. Participants*

Study participants were undergraduate students who were enrolled in computer science courses. A total of 227 students participated in the first session, and 145 in the second session. Study sessions were separated by one week. Study participants were compensated with course credit for participation. They were awarded credit for participation irrespective of their response. Participants were free to withdraw from the study at any point. Among the 101 participants, 75 were male, and 26 were female. This may create gender bias. Since this was a real classroom study and not a controlled experiment, gender distribution could not be controlled. Further, this represents the reality of gender distribution in higher education computer science classrooms. The average age of the participants was 20 years, with ages ranging from 18 to 27 years. There were 60 sophomores, 30 juniors, 10 seniors, and 1 freshman in the participant pool. There were 68 Asians, 32 White/Caucasians, 23 Hispanic/Latinos, 5 Black/African Americans, and 5 of two or more ethnicities.

## V. Data Analysis

The participants were filtered after data collection to obtain paired participants to ensure a within-subject study design for a methodologically sound analysis. After filtering the data and pairing participants, there were 101 participants. The data of these 101 participants were used for quantitative analysis.

The demographics and participants' prior experience data, and the count of responses are reported in the results section. For the quantitative analysis, all the data were gathered into an Excel sheet, and transformations were applied if required. IBM SPSS was used to conduct the statistical tests for the quantitative data. Parametric tests were used. If the data did not meet the normality test, then a nonparametric alternative was used. A threshold of 0.05 was used to determine significant differences when running a statistical test.

Student engagement as a factor scale [52] and student perspective of engagement in class scale (Assessing Student Perspective of Engagement in Class Tool: ASPECT) [53] were used as a questionnaire to measure students' engagement. The student engagement as a factor scale [52] was developed for the general classroom experience. It consisted of factors such as interest, enjoyment, challenge, skill, concentration, effort, importance, improved understanding, feeling in control, and goal clarity. The perspective of the engagement scale is a 16-item validated scale focused on assessing engagement in an active learning classroom. It included factors such as the Value of group activity (Value) and Personal Effort (PE) [53]. A Wilcoxon signed-rank test (a non-parametric test was used as the Shapiro-Wilk test revealed non-normal data distribution) was used to determine the difference in participants' self-reported scores on the questionnaire. The activity's usefulness in learning was measured using a single-item question "How much do you think the discussion activity (with peer/with ChatGPT) was useful in learning?" A Wilcoxon signed-rank test (a nonparametric test was used as the Shapiro-Wilk test revealed non-normal data distribution) was used to determine the difference in participants' self-reported scores for the activity's usefulness in learning. Further, to assess students' learning during activities, students were asked to report their level of learning at three stages of the study: (i) before reading the article, (ii) after reading the article, (iii) after the discussion activity (with a peer or AI tool). The Friedman test, a non-parametric alternative to repeated measures (used as the Shapiro-Wilk test revealed non-normal distribution), was used to determine any significant difference in self-reported knowledge between the two study conditions and three points of time, i.e., before reading the article, after reading the article, and after reading the article and having the discussion activity. A Wilcoxon signed-rank test (a nonparametric test was used as the Shapiro-Wilk test revealed non-normal data distribution) was used to assess any significant difference between learning test (a quiz consisting of 5 MCQ questions about the topic of the activity) scores.

Qualitative analysis was conducted on the peer discussion and AI-discussion data, i.e., the chat record. In the first session, 101 student conversations with ChatGPT were analyzed. Then, 16 chats were disregarded due to either invalid ChatGPT conversation links, repeated conversations, or missing links. This led to a total of 85 chats for analysis. The chat analysis sample was smaller than the quantitative analysis sample. The number of questions asked, feedback comments, follow-ups, and whether the student went off topic were identified by researchers by going through the conversation. All these factors were first defined by all the researchers together upon reviewing the data. The definitions below of all the measures were used for a qualitative analysis of the conversation. A question was something the participant asked in regard to the topic being discussed. The number of questions asked was determined by counting the no of questions posed by the participant to ChatGPT in the chat. A feedback comment is a response that evaluates an answer to a question. Off-topic is defined as a response that does not have any direct or indirect connection to the question asked. Follow-ups meant further pursuing the response, often by asking more questions to better understand the response to a question. The conversation begins with the students entering the given prompt and the article. This is shown in Fig. 3. Some students would ask ChatGPT to repeat the question or ask for a different question. For these conversations, ChatGPT may have a glitched response, prompting the student to ask for another question. In such situations, only questions that the student has answered were considered. A sample student conversation with ChatGPT is shown in Fig. 4. Further, the kinds of questions ChatGPT asked the students were analyzed. For the peer discussion condition (without AI condition), 82 conversations were used for analysis after filtering based on whether the chat was provided in participants' responses. Since this was a peer-to-peer discussion, the conversations consisted of a pair of participants, resulting in 54 unique conversations.

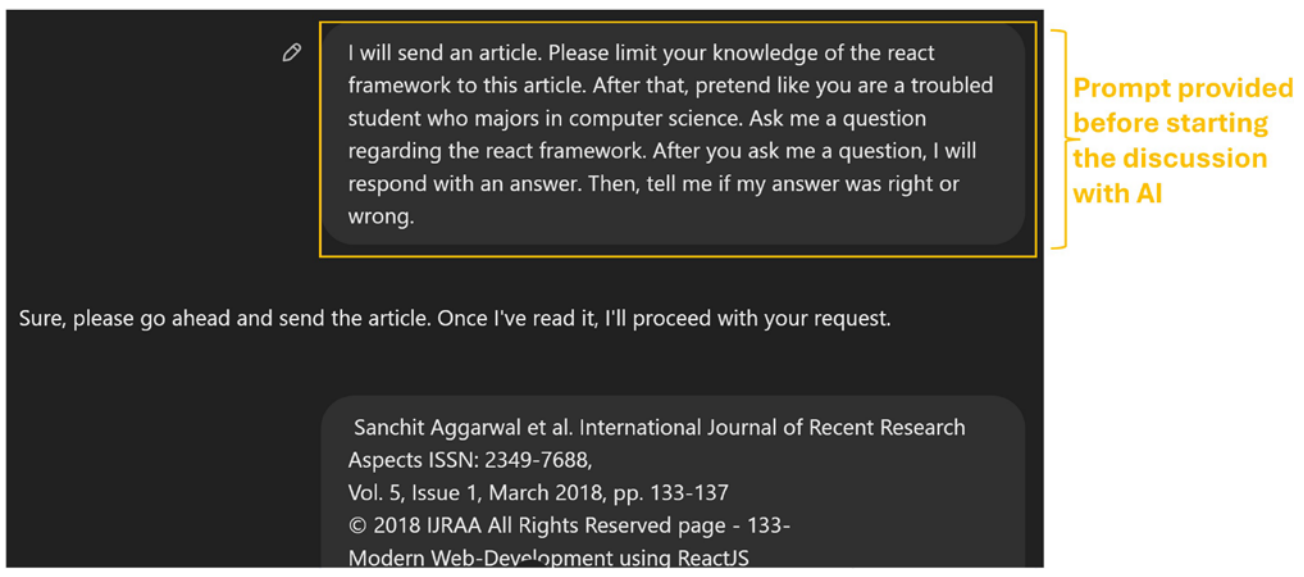

Fig. 3. The prompt each student copied and pasted before beginning the discussion with ChatGPT on the React article.

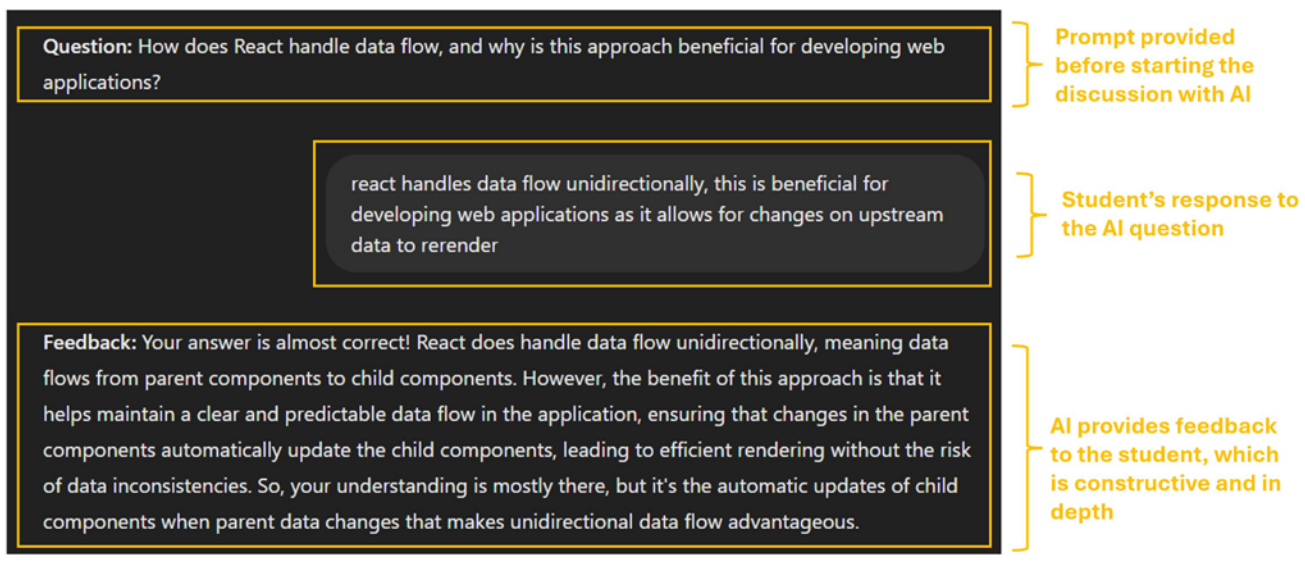

Fig. 4. Sample of conversation with ChatGPT during the discussion activity, where ChatGPT asks a question to the student, and the student responds. After which, ChatGPT provides feedback.

## VI. Results

In terms of the count of responses to the question asking if students had experience using an AI application, 89 students responded "yes", eight "maybe", and four "no". Students were also asked to rate their level of experience with AI on a scale of 1 to 10, 1 being no experience of using AI and 10 being a lot of experience of using AI. The average of the responses reported was 5.28. So, participants had above the median level of experience with AI. This can be attributed to the computing literacy of the students enrolled in computer science courses. Results of the Wilcoxon signed-rank test show a statistically significant difference in student engagement (student engagement as a factor scale) between the activity with the AI tool and the activity without the AI tool. Student engagement in activity with the AI tool (mean = 2.89, SD = 0.43) was rated significantly higher than activity without the AI tool (mean = 2.76, SD = 0.45); $z = -3.45$, $p < 0.001$. The effect size is calculated using $r = |z/N|$. A size effect of 0.45 was observed, indicating a medium to high size effect. The sub-constructs students' engagement as a factor scale, such as activity being interesting ($p < 0.001$), enjoyable ($p < 0.001$), excitement ($p = 0.004$), improved understanding ($p = 0.007$), challenging ($p$ = 0.034), and feeling in control ($p < 0.001$) were also significantly higher for activity with the AI tool compared to without the AI tool. The Wilcoxon signed-rank test for students' perspectives on engagement also illustrated a significant difference between the activity with the AI tool and the activity without the AI tool. Student engagement assessed through the student perspective of engagement in-class scale in activity with the AI tool (mean = 4.73, SD = 1.08) was rated significantly higher than activity without the AI tool (mean = 4.33, SD = 1.06); $z = -2.47$, $p = 0.014$. A small size effect with $r = 0.24$ was reported. For the activity's usefulness for learning, the Wilcoxon signed-rank test showed a difference between the activity with the AI tool and the activity without the AI tool. The activity with the AI tool (mean = 3.24, SD = 0.93) was rated significantly higher than the activity without the AI tool (mean = 2.72, SD = 0.69) for students' perceived usefulness of the activity in supporting learning; $z = -4.59$, $p < 0.001$.

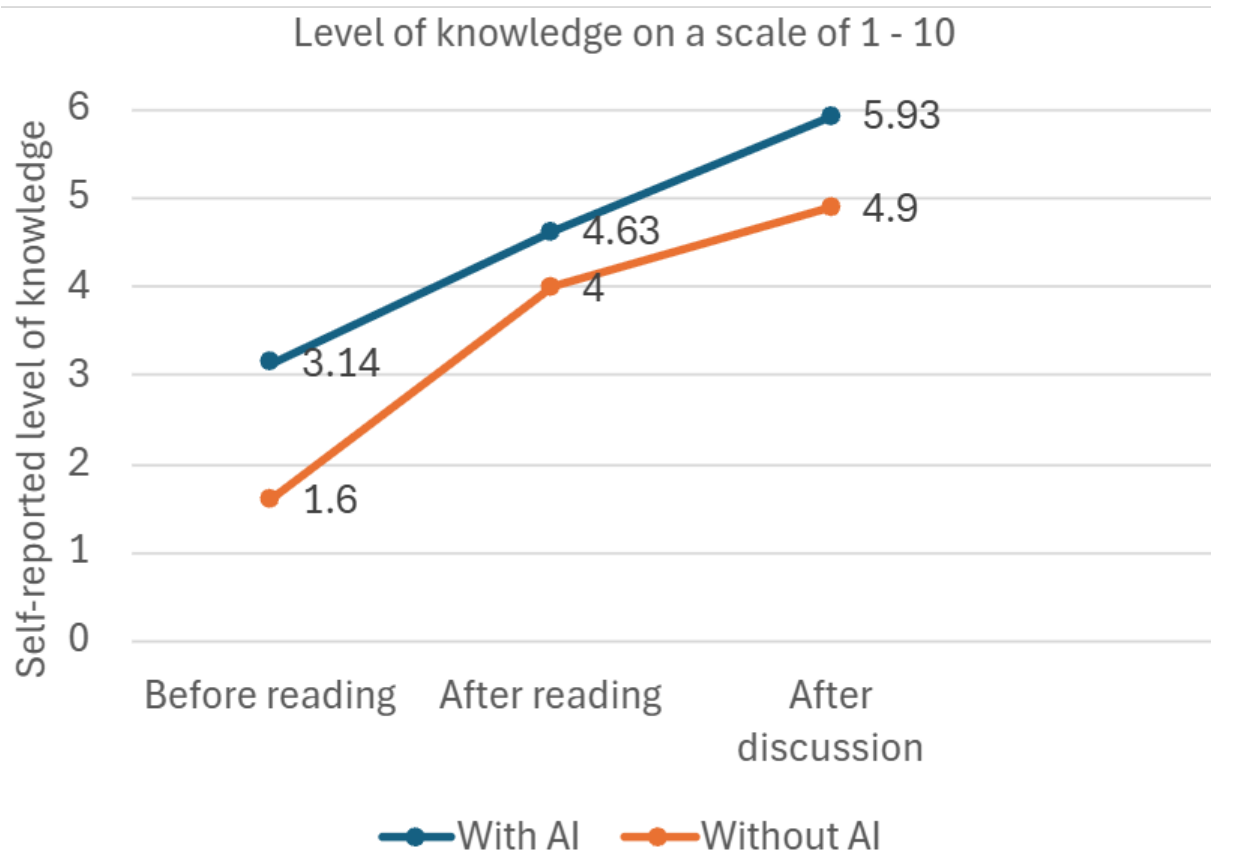

Fig. 5. Graph of the self-reported level of knowledge of reading material at three points in the study: before reading the article, after reading the article, and after discussion activity on the read article.

The results of the Friedman test on students' self-reported level of knowledge showed a statistically significant difference depending on stages in the learning activity (before reading, after reading, after reading and discussion), $\chi^2(5) = 334.78$, $p < 0.001$. The effect size is calculated using $w = \chi^2/N(k-1)$. A size effect of 0.66 was observed, indicating a large size effect. Fig. 5 shows the self-reported level of knowledge at three points of time, i.e., before reading the article, after reading the article, and after reading the article and having the discussion activity. Further, the Wilcoxon signed rank test for the test scores (learning outcome) at the end of each study condition also showed significant differences, $p = 0.015$, with the test scores being significantly higher for the AI condition (mean = 4.11, SD = 1.11) compared to the condition without AI (mean = 3.72, SD = 1.28). A small effect size ($r = 0.24$) was reported. Results have been summarized and presented in Fig. 6.

In the 85 students' conversations with ChatGPT analysis, we observed that for every answer entered by the student, ChatGPT evaluated the student's answer and gave feedback on it. The ChatGPT feedback typically began with stating if the answer was correct, partially correct, or incorrect, and then elaborating on what the correct answer is with details. A sample ChatGPT feedback on students' answers is shown in Fig. 4. On the contrary, the peer discussion was pretty short and not very descriptive, nor did it provide frequent feedback on responses, as can be seen in Fig. 7. In the 85 conversations analyzed, only once did the student go off topic. The average number of questions asked was 3.15. The number of questions asked ranged from 1 to 6. The average number of feedback comments was 3.12, ranging from 1 to 6. The mean number of follow-ups was 0.43, with the number of follow-ups ranging from 0 to 5.

| Summary of results | | | |
|---|---|---|---|
| Result | AI vs Peer discussion | Measure | p-value |
| Higher student engagement in the AI condition vs without | Discussion with AI> Discussion with Peer | Engagement as a factor scale | p<.001 |
| Higher student engagement in the AI condition vs without | Discussion with AI> Discussion with Peer | Student perspective of engagement scale | p=.014 |
| Higher usefulness for learning in the AI condition vs without | Discussion with AI> Discussion with Peer | Single item scale | p<.001 |
| Higher learning from activity in the AI condition vs without | Discussion with AI> Discussion with Peer | Single item scale | p<.001 |
| Higher test score in the AI condition vs without | Discussion with AI> Discussion with Peer | MCQ test score | p=.015 |
| More number of follow up during the discussion in the AI condition vs without | Discussion with AI> Discussion with Peer | Count of number of follow-ups | Avg no of follow-up (0.43>.33) |

Fig. 6. The summary of the results of the investigation.

Peer 1:
How familiar are you with AngularJS?

Peer 2:
I've heard of it before as another frontend framework like React.js but nothing further
what about you?

Peer 1:
Same here, more so just through word of mouth when people mention Javascript
How advantageous do you think SPAs are?

Peer 2:
I think it definitely depends on the project at hand but SPA seems to be very beneficial and easy to build
Do you think there are any setback of using AngularJS?

Peer 1:
Ah I see
It probably poses a problem for search engine optimization content using the structure is difficult to index
And that's particularly important as SEO becomes a bigger focus in years to come

Peer 1:
Apologies, let me ask
Do you think SPAs are overall better than static website pages?

Peer 2:
Overall, I do think SPA are better than static, because they allow for a dynamic website that wont have to rely on responses and re rendering the whole page and instead just key components

Fig. 7. A sample of peer discussion chat. As can be seen, the chat is to the point, and the responses are very short.

For the peer-to-peer discussion (without AI condition), the average number of questions asked was 3.25. The number of questions asked ranged from 1 to 7. The average number of feedback comments was 1.48, with the number of feedback comments ranging from 1 to 5. The average number of follow-ups was 0.33, with the number of follow-ups ranging from 0 to 3. There were 3 times the response was off-topic in the without AI condition (peer-to-peer discussion).

While the average number of questions asked is slightly higher in the without AI condition, it must be noted that responses were very short in peer discussion compared to ChatGPT. ChatGPT provided more comprehensive and more extended responses and feedback. Also, there was much more follow-up during the discussion with ChatGPT than with peers. In terms of response volume, the number of words in the discussion with AI condition was quite high compared to the conversations with peers in the second session. The verbose response primarily originated from ChatGPT, which is expected, as ChatGPT often generates verbose, human-like, meaningful responses in a conversation. Possibly, this is why students, when using AI for discussion, performed significantly better on the test and reported learning more compared to the peer discussion condition.

## VII. Discussion

The findings from this study provide indicative evidence of the significant role AI can play in enhancing understanding and engagement as opposed to not using AI. Specifically in large classroom settings, where there is often a lack of engagement between the students and the instructor, and the learning content. While we found that using CAI in class yielded positive results, it should be noted that this use was moderated and supervised by the instructor. Unsupervised use may not generate similar results. Further, the CAI interaction was somewhat structured by the starting prompt. Without the initial prompt, the interaction with CAI might result in some students taking the discussion in any direction they wanted or even directly asking for the response to the MCQ they were provided. Unstructured discussions may lead to varying levels of learning and discussion depth, depending on individual interests and curiosity. Some prior work has raised concerns about AI hallucinations and the accuracy of responses from ChatGPT. However, these concerns have been severely reduced with updated versions of these models. Nonetheless, there may be instances of partially erroneous output, and as AI models evolve, they will become more accurate.

The analysis of data from 101 paired participants reveals that those who utilized ChatGPT found the activity more interesting and enjoyable on average than those who did not use ChatGPT. Furthermore, the participants responded that the activity in which they used ChatGPT helped them learn better, improved their understanding, and was less challenging when compared with the responses from the activity without ChatGPT, as evident by the self-reported level of knowledge. The AI-generated personalized content tailored to student prompts and past responses, detailed explanations, and immediate feedback are likely the key factors contributing to these positive outcomes. Our results are in line with prior work by Yang and Ogata [54], where the use of a personalized learning environment resulted in improvements in both learning achievements and behavioral engagement for undergraduate students compared to traditional learning environments. The study was focused on blended learning in real class settings over a period of 6 weeks by providing actionable feedback to engage with the e-book. Prior work has also found evidence for AI tools supporting student engagement, while some prior work reported mixed results [41].

Our results also showed a significant difference in test scores, which is reflective of the use of AI positively impacting learning outcomes. Alternatively, we can say that the use of AI in an in-class activity resulted in higher engagement, which in turn also resulted in better learning outcomes and academic performance. This is in tandem with the prior work establishing a positive relationship between higher engagement and better learning outcomes [37].

The data also suggests that AI can transform the learning experience by making it more interactive and enjoyable, which likely contributes to a better understanding of the material. Participants who utilized AI tools likely felt more in control of their learning narrative and their pace. In prior work, the use of AI for personalized learning exercises also yielded positive results [26]. In both the study conditions, similar types of questions were asked, but conversations with AI tended to be more fruitful in terms of getting appropriate and elaborate responses. As can be seen in a sample peer discussion 5, responses are short and not very informative compared to AI responses. Almost a similar number of questions were asked with and without AI conditions. Nevertheless, participants in the AI condition score higher on the test (MCQ quiz). This could potentially mean that AI responses being longer, more in-depth, and explanatory in nature lead to higher learning despite asking a similar number of questions. Also, participants received feedback on their responses and additional information from the conversational AI. Also, there was more follow-up with AI than peers, possibly due to the expectation that AI knows much more and less fear of judgment for not knowing. Possibly due to interaction with the AI tool being private and not monitored, it makes students freer and empowered to ask any question that comes to mind and engage more freely. Such conversational, student-led inquiry-based learning has also been proposed in the literature to support student engagement [49]. This type of student-led interaction with AI may also support students' inquisitive nature and result in students developing a better understanding of the learning content. Our results also showed that the majority of participants, i.e., undergraduates enrolled in computer science courses, had some experience with AI applications, and they reported above the median level of experience with AI. This shows that AI can be applied to educational settings to support the learning process, aided by instructor guidance. While the CAI tool has the potential to support engagement and assist in learning, caution must be taken as it can also present challenges, such as plagiarism and overreliance on responses [12].

Although our results reveal promising potential for AI use in education, it must be noted that the results were reported in instructor-regulated AI use sessions. For the use of AI in formal education and its effective outcome, the presence and guidance of the instructors is crucial. It is worth noting that the study participants were computer science students, which may have contributed to the results and the efficient use of the AI tool. It was also observed that students went off topic more often in the condition without AI (peer-to-peer discussion) compared to the condition with AI. This can be attributed to a lack of human connection with AI. The shared social context between peers may have contributed to a higher incidence of off-topic conversations compared to the more task-focused interaction with the AI.

Further, the use of AI for discussion activity may result in better learning and more interaction, the number of questions asked, and follow-up, as ChatGPT is a conversational AI tool and often provides detailed responses. Also, AI conditions may have resulted in better learning outcomes as peers are likely to be limited in knowledge, while AI responds like an expert. AI responding like an expert may further motivate students to ask more and ask anything about the learning content of interest.

## VIII. Conclusion, Limitations, and Future Work

Our work explores AI for supporting engagement in a large

CS classroom. We explored the use of AI for an in-class reading activity, followed by a discussion with AI. We conducted a quasi-experimental within-subjects study comparing the use of AI in an in-class activity to an in-class activity without AI (with a peer). In response to RQ1, our study found that AI has the potential to enhance engagement with learning content. AI tends to play the role of a learned agent that provides students with explanations for their queries. This opens up potential ways AI could be used to increase engagement in large classrooms and formal learning settings. In response to RQ2, our study's results show that students perceive AI as helpful for learning. Both the self-reported AI usefulness score and the post-task quiz score support the finding of higher perceived usefulness of the learning activity with AI than without. Further, the analysis of AI and peer interactions showed that students tend to follow up more when interacting with AI. These positive results of AI use can be explained by the role of AI as a learned agent, leading students to ask more of AI than they would of a peer who has a similar level of knowledge as them. A higher quiz score can be supported by the quality of AI responses, with in-depth explanations and human-like language, which make the interaction more useful for students' learning.

Overall, our research indicated that AI has a positive impact on formal education settings. However, there were some limitations, such as the results being based on a single AI use session. Also, the study conditions were not counterbalanced due to real classroom settings, which prohibited differential treatment of the course population. Furthermore, the study session was monitored and regulated by the instructor. Without higher authority guidance, AI usage may change. The study was conducted in a hybrid CS course at a United States university, which may limit the generalizability of results. Further, our study primarily used self-reported responses, supplemented by some objective data. Nonetheless, our study results and contributions remain valuable to researchers and educators, as the study was conducted in real classroom settings and offers authentic AI interaction in formal settings.

Building on this research, future work should conduct longitudinal studies to understand the impact of long-term AI use on engagement with learning content. This will validate our study and illustrate sustained engagement. Additionally, to include AI as part of learning in a formal setting, exploration is needed to understand its impact on learning outcomes and on more tangible cognitive skill development, such as its effects on students' cognitive abilities and critical thinking skills, using objective data. Future research should explore the use of AI to support education beyond university students, including the K-12 population. Future work should explore the use of AI in remote learning to enable personalized learning experiences.

## Conflict of Interest

The authors declare no conflict of interest

## Author Contributions

Neha Rani, Sharan Majumder, and Ishan Bhardwaj ideated the project, brainstormed study details, and conducted the research; Neha Rani, Sharan Majumder, Ishan Bhardwaj, and Pedro Guillermo Feijoo Garcia conducted the qualitative and quantitative data analysis; Neha Rani, Sharan Majumder, and Ishan Bhardwaj drafted the manuscript. All authors have reviewed and approved the final version.

## Acknowledgement

We thank the participants of the study for their time. We also acknowledge the use of an AI tool for grammar and language enhancement and reference management.

## References

[1] P. Denny, J. Prather, B. A. Becker, J. Finnie-Ansley, A. Hellas, J. Leinonen, A. Luxton-Reilly, B. N. Reeves, E. A. Santos, and S. Sarsa, "Computing education in the era of generative AI," *Communications of the ACM*, vol. 67, no. 2, pp. 56–67, 2024.

[2] C. Diwan, S. Srinivasa, G. Suri, S. Agarwal, and P. Ram, "AI-based learning content generation and learning pathway augmentation to increase learner engagement," *Computers and Education: Artificial Intelligence*, vol. 4, 100110, 2023.

[3] F. Tahiru, "Ai in education: A systematic literature review," *Journal of Cases on Information Technology (JCIT)*, vol. 23, no. 1, pp. 1–20, 2021.

[4] X. Zhai, X. Chu, C. S. Chai, M. S. Jong, A. Istenic, M. Spector, J. B. Liu, J. Yuan, and Y. Li, "A review of Artificial Intelligence (AI) in education from 2010 to 2020," *Complexity*, vol. 2021, no. 1, 8812542, 2021.

[5] M. Chassignol, A. Khoroshavin, A. Klimova, and A. Bilyatdinova, "Artificial intelligence trends in education: A narrative overview," *Procedia Computer Science*, vol. 136, pp. 16–24, 2018.

[6] J. Prather, P. Denny, J. Leinonen, B.A. Becker, I. Albluwi, M. Craig, H. Keuning, N. Kiesler, T. Kohn, A. Luxton-Reilly, and S. MacNeil, "The robots are here: Navigating the generative ai revolution in computing education," in *Proc. the 2023 Working Group Reports on Innovation and Technology in Computer Science Education*, 2023, pp. 108–159.

[7] C. Bird, D. Ford, T. Zimmermann, N. Forsgren, E. Kalliamvakou, T. Lowdermilk, and I. Gazit, "Taking flight with copilot: Early insights and opportunities of AI-powered pair-programming tools," *Queue*, vol. 20, no. 6, pp. 35–57, 2022.

[8] R. Liu, C. Zenke, C. Liu, A. Holmes, P. Thornton, and D. J. Malan, "Teaching CS50 with AI: Leveraging generative artificial intelligence in computer science education," in *Proc. the 55th ACM Technical Symposium on Computer Science Education V. 1*, 2024, pp. 750–756.

[9] P. Denny, B. A. Becker, J. Leinonen, and J. Prather, "Chat overflow: Artificially intelligent models for computing education-renaissance or apocaiypse?" in *Proc. the 2023 Conference on Innovation and Technology in Computer Science Education V. 1*, 2023, pp. 3–4.

[10] H. Oubalahcen, L. Tamym *et al.*, "The use of ai in e-learning recommender systems: A comprehensive survey," *Procedia Computer Science*, vol. 224, pp. 437–442, 2023.

[11] S. Munawar, S. K. Toor, M. Aslam, and M. Hamid, "Move to smart learning environment: Exploratory research of challenges in computer laboratory and design intelligent virtual laboratory for elearning technology," *EURASIA Journal of Mathematics, Science and Technology Education*, vol. 14, no. 5, pp. 1645–1662, 2018.

[12] C. K. Lo, K. F. Hew, and M. S.-Y. Jong, "The influence of ChatGPT on student engagement: A systematic review and future research agenda," *Computers & Education*, 105100, 2024.

[13] J. G. Cullen, "The writing skills course as an introduction to critical practice for larger business undergraduate classes," *International Journal of Management Education*, 2011.

[14] C. Mulryan-Kyne, "Teaching large classes at college and university level: Challenges and opportunities," *Teaching in Higher Education*, vol. 15, no. 2, pp. 175–185, 2010.

[15] F. Iniaghe and C. Osiobe, "Students' perception of class size as an indicator of academic performance of senior secondary school students in uvwie local government area of delta state, Nigeria," *International Journal of Trends and Developments in Education*, vol. 4, no. 1, pp. 59–73, 2024.

[16] T. L. Wentling, J. Park, and C. Peiper, "Learning gains associated with annotation and communication software designed for large undergraduate classes," *Journal of Computer Assisted Learning*, vol. 23, no. 1, pp. 36–46, 2007.

[17] D. Allen and K. Tanner, "Infusing active learning into the large enrollment biology class: Seven strategies, from the simple to complex," *Cell Biology Education*, vol. 4, no. 4, pp. 262–268, 2005.
[18] A. Kothiyal, S. Murthy, and S. Iyer, "Think-pair-share in a large CS1 class: Does learning really happen?" in *Proc. the 2014 Conference on Innovation & Technology in Computer Science Education*, 2014, pp. 51–56.
[19] M. Mazzara, P. Zhdanov, M. R. Bahrami, H. Aslam, I. Kotorov, M. Imam, H. Salem, J. A. Brown, and R. Pletnev, "Education after COVID-19," in *Smart and Sustainable Technology for Resilient Cities and Communities*, Springer, 2022, pp. 193–207.
[20] K. L. Hourigan, "Increasing student engagement in large classes: The arc model of application, response, and collaboration," *Teaching Sociology*, vol. 41, no. 4, pp. 353–359, 2013.
[21] R. Goff, J. Terpenny, and T. Wildman, "Improving learning and engagement for students in large classes," in *Proc. 2007 37th Annual Frontiers in Education Conference on Global Engineering: Knowledge Without Borders, Opportunities Without Passports*, IEEE, 2007.
[22] G. Nishimwe, S. Kamali, E. Gatesi, and R. Wong, "Assessing the perceptions and preferences between online and in-person classroom learning among university students in Rwanda," *Journal of Service Science and Management*, vol. 15, no. 1, pp. 23–34, 2022.
[23] M. Kazemitabaar, R. Ye, X. Wang, A.Z. Henley, P. Denny, M. Craig, and T. Grossman, "CodeAID: Evaluating a classroom deployment of an llm-based programming assistant that balances student and educator needs," in *Proc. the CHI Conference on Human Factors in Computing Systems*, 2024, pp. 1–20.
[24] M. Liffiton, B. E. Sheese, J. Savelka, and P. Denny, "Codehelp: Using large language models with guardrails for scalable support in programming classes," in *Proc. the 23rd Koli Calling International Conference on Computing Education Research*, 2023, pp. 1–11.
[25] S. Sarsa, P. Denny, A. Hellas, and J. Leinonen, "Automatic generation of programming exercises and code explanations using large language models," in *Proc. the 2022 ACM Conference on International Computing Education Research-Volume 1*, 2022, pp. 27–43.
[26] E. Logacheva, A. Hellas, J. Prather, S. Sarsa, and J. Leinonen, "Evaluating contextually personalized programming exercises created with generative AI," in *Proc. the 2024 ACM Conference on International Computing Education Research-Volume 1*, 2024, pp. 95–113.
[27] B. Kim, H. Suh, J. Heo, and Y. Choi, "Ai-driven interface design for intelligent tutoring system improves student engagement," arXiv Preprint, arXiv:2009.08976, 2020.
[28] C. H. Wu and V. T. Ho, "Critical factors for why ChatGPT enhances learning engagement and outcomes," *Education and Information Technologies*, pp. 1–32, 2025.
[29] D. T. K. Ng, C. W. Tan, and J. K. L. Leung, "Empowering student self-regulated learning and science education through ChatGPT: A pioneering pilot study," *British Journal of Educational Technology*, vol. 55, no. 4, pp. 1328–1353, 2024.
[30] M. Yang, S. Jiang, B. Li, K. Herman, T. Luo, S. C. Moots, and N. Lovett, "Analysing nontraditional students' ChatGPT interaction, engagement, self-efficacy and performance: A mixed-methods approach," *British Journal of Educational Technology*, 2025.
[31] A. Rissanen, "Student engagement in large classroom: The effect on grades, attendance and student experiences in an undergraduate biology course," *Canadian Journal of Science, Mathematics and Technology Education*, vol. 18, pp. 136–153, 2018.
[32] B. Macfarlane and M. Tomlinson, "Critiques of student engagement," *Higher Education Policy*, vol. 30, pp. 5–21, 2017.
[33] J. A. Fredricks, P. C. Blumenfeld, and A. H. Paris, "School engagement: Potential of the concept, state of the evidence," *Review of Educational Research*, vol. 74, no. 1, pp. 59–109, 2004.
[34] G. Heaslip, P. Donovan, and J. G. Cullen, "Student response systems and learner engagement in large classes," *Active Learning in Higher Education*, vol. 15, no. 1, pp. 11–24, 2014.
[35] G. Iaria and H. Hubball, "Assessing student engagement in small and large classes," *Transformative Dialogues: Teaching and Learning Journal*, vol. 2, no. 1, 2008.
[36] A. Usher and A. Cervenan, "Global higher education rankings: Affordability and accessibility in comparative perspective, 2005," *Online Submission*, 2005.
[37] R. M. Carini, G. D. Kuh, and S. P. Klein, "Student engagement and student learning: Testing the linkages," *Research in Higher Education*, vol. 47, pp. 1–32, 2006.
[38] P. T. Ewell, "The us National Survey of Student Engagement (NSSE)," *Public Policy for Academic Quality: Analyses of Innovative Policy Instruments*, pp. 83–97, 2010.
[39] A. Ikedinachi, S. Misra, P. A. Assibong, E. F. Olu-Owolabi, R. Maskeliunas, and R. Damasevicius, "Artificial intelligence, smart classrooms and online education in the 21st century: Implications for human development," *Journal of Cases on Information Technology (JCIT)*, vol. 21, no. 3, pp. 66–79, 2019.
[40] A. Harry and S. Sayudin, "Role of AI in education," *Interdiciplinary Journal and Hummanity (INJURITY)*, vol. 2, no. 3, pp. 260–268, 2023.
[41] Y. M. E. Heung and T. K. Chiu, "How ChatGPT impacts student engagement from a systematic review and meta-analysis study," *Computers and Education: Artificial Intelligence*, vol. 8, 100361, 2025.
[42] P. Diwanji, K. Hinkelmann, and H. F. Witschel, "Enhance classroom preparation for flipped classroom using ai and analytics," in *Proc. the 20th International Conference on Enterprise Information Systems (ICEIS 2018)*, 2018, pp. 477–483.
[43] R. Winkler and J. Roos, "Bringing AI into the classroom: Designing smart personal assistants as learning tutors," in *Proc. the Fortieth International Conference on Information Systems*, 2019.
[44] Y. Chen, S. Jensen, L. J. Albert, S. Gupta, and T. Lee, "Artificial Intelligence (AI) student assistants in the classroom: Designing chatbots to support student success," *Information Systems Frontiers*, vol. 25, no. 1, pp. 161–182, 2023.
[45] W. Alharbi, "AI in the foreign language classroom: A pedagogical overview of automated writing assistance tools," *Education Research International*, vol. 2023, no. 1, 4253331, 2023.
[46] W. Inman. (Jan. 2024). National institute for student success at Georgia state awarded $7.6m to study benefits of ai-enhanced classroom chatbots. Georgia State News Hub. [Online]. Available: https://news.gsu.edu/2024/01/11/nationalinstitute-for-student-success-awarded-7-6-million-grant-by-us-department-of-education/
[47] Y.-F. Lee, G.-J. Hwang, and P.-Y. Chen, "Impacts of an ai-based cha bot on college students' after-class review, academic performance, self-efficacy, learning attitude, and motivation," *Educational Technology Research and Development*, vol. 70, no. 5, pp. 1843–1865, 2022.
[48] M. N. Hadian, E. Wihardjo, I. Murfidah, and E. S. W. Jannah, *Analysis of the Impact of Artificial Intelligence and Gamification in Enhancing Student Engagement in Virtual Learning Environments*, 2024.
[49] M. A. Adarkwah, C. Ying, M. Y. Mustafa, and R. Huang, "Prediction of learner information-seeking behavior and classroom engagement in the advent of ChatGPT," in *Proc. International Conference on Smart Learning Environments*, Springer, 2023, pp. 117–126.
[50] G. Pitts, N. Rani, W. Mildort, and E.-M. Cook, "Students' reliance on ai in higher education: Identifying contributing factors," arXiv Preprint, arXiv:2506.13845, 2025.
[51] G. R. Bai, Z. Jiang, T. W. Price, and K. T. Stolee, "Evaluating the effectiveness of a testing checklist intervention in CS2: An quasi-experimental replication study," in *Proc. the 2024 ACM Conference on International Computing Education Research Volume 1*, 2024, pp. 55–64.
[52] D. J. Shernof, E. A. Ruzek, A. J. Sannella, R. Y. Schorr, L. Sanchez Wall, and D. M. Bressler, "Student engagement as a general factor of classroom experience: Associations with student practices and educational outcomes in a university gateway course," *Frontiers in psychology*, vol. 8, 994, 2017.
[53] B. L. Wiggins, S. L. Eddy, L. Wener-Fligner, K. Freisem, D. Z. Grunspan, E. J. Theobald, J. Timbrook, and A. J. Crowe, "Aspect: A survey to assess student perspective of engagement in an active learning classroom," *CBE—Life Sciences Education*, vol. 16, no. 2, 32, 2017.
[54] C. C. Yang and H. Ogata, "Personalized learning analytics intervention approach for enhancing student learning achievement and behavioral engagement in blended learning," *Education and Information Technologies*, vol. 28, no. 3, pp. 2509–2528, 2023.